# Restricted Structural Random Matrix for Compressive Sensing

Thuong Nguyen Canh and Byeungwoo Jeon*

*Department of Electronic, Electrical and Computer Engineering, Sungkyunkwan University*

**Abstract**

Compressive sensing (CS) is well-known for its unique functionalities of sensing, compressing, and security (i.e. CS measurements are equally important). However, there is a tradeoff. Improving sensing and compressing efficiency with prior signal information tends to favor particular measurements, thus decrease the security. This work aimed to improve the sensing and compressing efficiency without compromise the security with a novel sampling matrix, named Restricted Structural Random Matrix (RSRM). RSRM unified the advantages of frame-based and block-based sensing together with the global smoothness prior (i.e. low-resolution signals are highly correlated). RSRM acquired compressive measurements with random projection (equally important) of multiple randomly sub-sampled signals, which was restricted to be the low-resolution signals (equal in energy), thereby, its observations are equally important. RSRM was proven to satisfies the Restricted Isometry Property and shows comparable reconstruction performance with recent state-of-the-art compressive sensing and deep learning-based methods.

## 1. Introduction

Compressive Sensing (CS) [1, 2, 3, 4] is an emerging sampling technique that directly captures a sparse or compressible signal $x \in \mathbb{R}^n$ in a compressed form $y \in \mathbb{R}^m, m \ll n$ via random projection:

$$y = \Phi x, \qquad (1)$$

where $\Phi \in \mathbb{R}^{m \times n}$ denotes the sampling matrix. CS is well known for its unique functionalities of simultaneous sampling, compressing, and security (or democracy).

**Sensing**. With the linear projection sampling, CS guarantees to reconstruct signal at a high probability if the sampling matrix $\Phi$ satisfies some condition, like the Restricted Isometry Property (RIP) [3, 4], as

*Definition 1*. *A matrix $\Phi$ is said to satisfy the RIP condition with an isometric constant $\delta \in (0,1)$ if the inequality*

$$(1 - \delta)\|\alpha\|_2^2 \leq \|\Phi\alpha\|_2^2 \leq (1 + \delta)\|\alpha\|_2^2, \qquad (2)$$

*is held for all $s$-sparse signals $\alpha \in \mathbb{R}^n$. The smaller $\delta$ becomes, the better RIP condition.*

The asymmetric RIP condition was developed [53, 54] to allow the difference in upper and lower bound.

---

\* Corresponding author. *E-mail address:* bjeon@skku.edu.

# Thuong Nguyen Canh was with College of Information and Communication Engineering, Sungkyunkwan University, South Korea. He is now with Institute for Datability Science, Osaka University, Japan.

# Byeungwoo Jeon is with College of Information and Communication Engineering, Sungkyunkwan University, South Korea.

In the literature, CS widely uses the Gaussian Random Matrix (GRM) due to its theoretical guarantee. Unfortunately, GRM requires *significant computation* and *storage* for its fully random structure. Over the past decades, researchers have sought to reduce its computational complexity. These studies can be classified as (i) block-based CS (BCS), which relies on processing signals in a block-based manner [5, 6, 7]; (ii) frame-based sensing, which samples each signal dimension separately in Kronecker CS (KCS) [8-11], random sample signals in fast-transform domains [12, 13], circulant random matrix [14], or sparse random matrices [15]. In general, frame-based CS is preferred for high sampling efficiency [8-13] and BCS for coding efficiency [16, 18, 19], parallel sampling [37].

**Compressing**. As the number of acquired measurements is significantly smaller than the signal dimension, $m \ll n$, CS achieves compressing capabilities. As a sampling technique, CS assumes no information about the to-be-sampled signal, except for the sparsity prior. Meanwhile, standard compression methods (e.g., JPEG and HEVC) know the to-be-compressed signals, thereby encode them at higher coding efficiencies than CS [16, 17]. To overcome this drawback, researchers have been *utilized other signal priors*, such as the low-frequency prior (i.e. the human visual system is more sensitive to lower frequencies) [20-26], structural of signals (i.e. tree-sparse in wavelet [27] or zig-zag order in DCT [28]), perceptual prior (in image/video applications [29]), and smoothness prior (nearby samples are highly correlated) [16-19]. Multi-scale CS [20-24] and hybrid CS [25, 26] captured more low-frequency components to utilize the low-frequency prior. Weighted sampling exploited the structure of signal [27-29] to capture more important components without influencing the perceptual quality. The smoothness prior [16-19] was used for predictive coding in BCS because the nearby block measurements are highly correlated.

**Security**. CS measurements are equally important due to the randomness of the sampling matrix, thus named democracy property [30]. It is possible to recover signals from CS acquisitions despite damaged/degraded or lost measurements, which enable error-resilient applications [31]. Additionally, if GRM is used as a sampling matrix, then CS measurements are also following the Gaussian distribution. Therefore, despite revealing the energy of the signal, the sampling matrix was used as the private key providing a certain level of security [32-36]. As a result, numerous studies have attempted to analyze the theoretical security performance [32, 33], practical applications in multimedia capturing [50, 51] and/or transmission including cloud computing (e.g., for error resilience [31], to improve visual privacy protection [34], or for encryption [35, 36], cloud server-side protection [50, 51]). CS was further combined with conventional encryption techniques like random scrambling/flipping, chaos methods [51], etc.

There is, however, a trade-off between these functionalities. Researchers often used prior information to reduce sampling computation or improve sampling efficiency without considering the security. Utilizing prior information leads to reducing the randomness of the projection, tends to favour particular measurements. As a result, CS measurements are no longer equally important. For instance, reducing complexity with BCS reveals the signal structure by evaluating the block measurement's energy. Alternatively, CS-based security methods reduce sampling and compressing performance and may even increase the sampling complexity.

In this work, we develop Generalized Structural Sensing Matrix (GSSM) and derived a special version named Restricted Structural Random Matrix (RSRM). RSRM is targeted to improve sensing efficiency, reduce complexity while maintaining the security/democracy property. *Theoretically*, RSRM unifies the advantages of BCS, the structural frame-based, and utilizes the smoothness prior which is held for compressible and structure sparse signals. We prove the asymmetric RIP condition of GSSM, symmetric RIP condition of RSRM together with RIP supported experiments. *Practically*, RSRM is a combination of CS with conventional under-sampling methods such as partial sub-sampling, multiple-images super-resolution, and coded imaging. RSRM randomly sub-samples multiple LR signals, which is equal in energy, then apply random block-based sampling for each. Therefore, each measurement of RSRM is equally important, thus the democracy is preserved. Note that, our work focusses on the security approach in the encoder part in which security, sampling, and compression are performed simultaneously.

The contributions of this paper are summarized are following
- Propose a Generalized Structural Sensing Matrix from block-based and structural frame-based sampling.
- Present the asymmetric RIP of GSSM and derived the conditions to improve its RIP condition.
- Propose a Restricted Structural Random Matrix as a special case of GSSM by utilizing the local smoothness prior via global manner with Restricted Permutation.

We briefly review the related works in Section II. Section III introduces the generalized structural sensing matrix and its asymmetric RIP condition. Section IV proposes the restricted structure random matrix, a special version of GSSM, to achieve the best RIP condition taking the advantage of the global smoothness prior, especially for compressible and structure sparse signals. We present support experiments and reconstruction performance in Section V. Finally, we draw conclusions in Section VI.

For notation convention, a scalar value is written in normal form as $m$, a vector is represented in normal bold as $\boldsymbol{x}$, while a matrix is denoted in capital bold notation as $\boldsymbol{X}$. Other notations are given in Table 1.

Table 1. Notations and descriptions

| Notation | Descriptions | Notation | Descriptions |
|---|---|---|---|
| $\boldsymbol{x}$ | 1D signal | $\boldsymbol{y}$ | 1D measurement |
| $\boldsymbol{X}$ | 2D signal | $\boldsymbol{Y}$ | 2D measurement |
| $\boldsymbol{\Phi}_i$ | 2D sampling matrix of GRM | $\boldsymbol{\Phi}_{METHOD}$ | 2D sampling matrix of METHOD |
| $\boldsymbol{\Phi}_i^B$ | 2D random matrix GRM, full rank | $\boldsymbol{R}_i, \boldsymbol{D}_i$ | 2D Partially sub-sampling matrix |
| $\boldsymbol{T}_i$ | 2D transform matrix like DCT | $n$ | 1D signal dimension |
| $n_B$ | 1D block size | $m = \sum m_i$ | Scalar, number of measurements |
| $c$ | Scalar, number of sub-sampled signal | $m_i$ | Scalar, number of measurements at block $i$ |

## 2. Related Work

### 2.1. Sampling Complexity Reduction

#### 2.1.1. Block-Based Compressive Sensing (BCS)

Motivated by the block-by-block processing of the JPEG and MPEG standards, block-based CS (BCS) [5, 6] divides $n$-d signal to $c$ non-overlapping blocks of size $n_B$, then samples each with the sampling matrix $\boldsymbol{\Phi}_i$ as

$$\boldsymbol{\Phi}_{BCS} = \begin{bmatrix} \boldsymbol{\Phi}_1 & 0 & 0 & 0 \\ 0 & \boldsymbol{\Phi}_2 & 0 & 0 \\ \ldots & \ldots & \ddots & \vdots \\ 0 & 0 & 0 & \boldsymbol{\Phi}_c \end{bmatrix}, \quad (3)$$

where $m = \sum_i m_i$, $n = c * n_B$ and $\boldsymbol{\Phi}_i \in \mathbb{R}^{m_i \times n_B}, m_i \ll n_B$, is a GRM matrix. As a result, BCS demands less memory storage and computational complexity for the sampling process as well as reconstruction (i.e., reconstruction requires multiple matrix multiplications with the sampling matrix). Each BCS observation carries information about a nonoverlap portion of the signal, thereby, enables parallel sampling and reconstruction. We can reduce the acquisition time with the single pixel camera [37]. However, the energy of each block measurement might reveal a low-resolution version of the signals as in [38], thereby BCS compromises the democracy property.

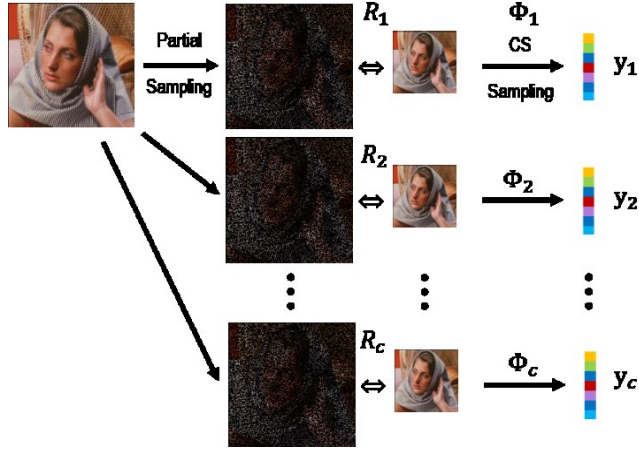

Fig. 1. Proposed RSRM: random sample of $c$ sub-sampled images which are restricted to be low-resolution images.

### 2.1.2. Frame-Based Sensing: A Structurally Random Matrix

The scrambled Fourier transform [3] was the first structurally random matric by randomly selecting Fourier coefficients primally used for MRI acquisition. It was further generalized by T. T. Do et al. [12] to the structurally random matrix (SRM). SRM first permutates the signal, then captures measurements by scrambled transform matrix $T \in \mathbb{R}^{n \times n}$ which is created by randomly selects $m$ rows of a fast transform matrix (e.g., DCT) as

$$\Phi_{SRM} = DTR, \qquad (4)$$

where $D \in \mathbb{B}^{m \times n}$ and $R \in \mathbb{B}^{n \times n}$ denote the random sub-sampling and permutation matrices, respectively. A scale factor $\sqrt{n/m}$ is often used to normalize the energy of measurements. For efficiency sampling, a block-based version of SRM (BSRM) is also considered as

$$\Phi_{SRM} = \begin{bmatrix} D_1 & 0 & 0 & 0 \\ 0 & D_2 & 0 & 0 \\ \cdots & \cdots & \ddots & \vdots \\ 0 & 0 & 0 & D_c \end{bmatrix} \begin{bmatrix} T_1 & 0 & 0 & 0 \\ 0 & T_2 & 0 & 0 \\ \cdots & \cdots & \ddots & \vdots \\ 0 & 0 & 0 & T_c \end{bmatrix} \begin{bmatrix} R_1 \\ R_2 \\ \vdots \\ R_c \end{bmatrix} \qquad (5)$$

Since the sampling order is not important in CS, we reorder the elements of $D$ so that it becomes a block diagonal of a smaller size $D_i \in \mathbb{B}^{m_i \times n_B}$. The transform matrix is also a block diagonal matrix at a smaller size of $T_i = T_B \in \mathbb{R}^{n_B \times n_B}$. The permutation matrix $R$ can be divided horizontally into $c$ sub-sampling matrices $R_i \in \mathbb{B}^{n_B \times n}$. Note that, each row and column of $D$ and $R$ have only one position of value 1. SRM enables fast sampling and reconstruction and can be interpreted as BCS of a permutated signal. Unlike BCS, each block signal is randomly sub-sampled from the original signal; thus, they are equally important. Even though BSRM reveals no information about the signal via evaluating measurement, using a well-known fast transform significantly reduces the randomness of the sampling matrix. Since the number of sparsifying transform is limited, the attacker can guess the sampling matrix faster. Thus, this work utilizes the block-based GRM random matrix instead. Using GRM adds more complexity to the conventional BSRM but still much faster than fully random matrix with GRM while outperforming reconstruction performance and preserve the democracy. Other frame-based matrices like binary, circulant matrices, chaos method, etc. are out of the scope of this paper.

## 2.2. Conventional Under-Sampling Methods

Prior to CS, conventional methods to reduce the sampling rate are partial sampling (PS), super-resolution (SR) from single/multiple images, and coded imaging (CI). In PS, a signal is acquired at random locations with or without well-designed patterns. Instead of sampling a high-resolution (HR) signal, low-resolution (LR) signals are captured and recovered missing high-frequency contents by super-resolution. Coded imaging was designed to increase the frame-rate by assigning each frame a coded mask.

At a low subrate, CS also performs poorly, even with a GRM. Therefore, it is straightforward to combine conventional under-sampling methods and CS to further improve sampling efficiency. For CS+PS, the authors [44, 45] captured CS measurements form the partially sampled signal. For CS+SR, the authors [46, 47] sampled multiple LR measurements and performed super resolution-based reconstruction. CS has been combined with CI to improve the temporal imaging in [48]. Our work combines all methods (PS, SR, and CI) into CS, as illustrated in Fig. 1. Firstly, we utilize PS, SR, and CI by capturing multiples LR images with unique random sub-sampling patterns. We, then, acquire CS measurements of each random LR images. Thus, reconstruction from our proposed measurements is equivalent to recovering images from multiple compressed LR measurements.

## 3. Generalized Sparse Structural Matrix

### 3.1. Proposed Generalized Sparse Structural Matrix (GSSM)

From the definition of BCS and BSRM, we proposed a generalized sparse structural matrix (GSSM) as

$$\Phi_{GSSM} = \begin{bmatrix} D_1 & 0 & 0 & 0 \\ 0 & D_2 & 0 & 0 \\ \dots & \dots & \ddots & \vdots \\ 0 & 0 & 0 & D_c \end{bmatrix} \begin{bmatrix} \Phi_1^B & 0 & 0 & 0 \\ 0 & \Phi_2^B & 0 & 0 \\ \dots & \dots & \ddots & \vdots \\ 0 & 0 & 0 & \Phi_c^B \end{bmatrix} \begin{bmatrix} R_1 \\ R_2 \\ \vdots \\ R_c \end{bmatrix},$$
$$\Leftrightarrow \Phi_{GSSM} = D\Phi^B R, \qquad \Phi_i = D_i \Phi_i^B$$

(6)

- $R \in \mathbb{B}^{cn_B \times n}$ is constructed by concatenating $c$ sub-sampling matrices $R_i \in \mathbb{B}^{n_B \times n}$ which is generated by selecting $n_B$ random subsets of the identity matrix $I_n \in \mathbb{B}^{n \times n}$. Let $p$ and $q$ denote the minimum and maximum number of times a sample in $c$ sub-sampling matrices $R_i$, $0 < p \le q \le c$. We stress that, in this paper, we do not consider the extreme case of $p = 0$ since it throws away the signal information.
- $\Phi \in \mathbb{R}^{cn_B \times cn_B}$ is a block diagonal matrix with $c$ sensing matrices $\Phi_i^B \in \mathbb{R}^{n_B \times n_B}$.
- $D \in \mathbb{B}^{m \times cn_B}$ is a block diagonal matrix with $c$ sub-sampling matrices $D_i \in \mathbb{R}^{m_i \times n_B}$ which is generated by selecting $m_i$ random subsets of the identity matrix $I_{n_B} \in \mathbb{B}^{n_B \times n_B}$.

We interpret GSSM as a two-step sampling of: (i) sub-sample the to-be-sampled signal $c$ times, $x_i = R_i x$, then (ii) capture each sub-sampled signal $x_i$ with a corresponding matrix $D_i \Phi_i^B$. GSSM measurements can be formulated as

$$y = [y_1; y_2; \cdots; y_c], \qquad y_i = D_i \Phi_i^B R_i x. \qquad (7)$$

From its definition, GSSM is a generalization of BSRM and BCS with extensions. Firstly, GSSM relaxes the condition $cn_B = n$ in BCS and BSRM to $cn_B \ge n$ so that $R$ can be a non-square matrix. Secondly, $\Phi_i^B$ can be any sampling matrix and is not limited to a fast transform as in BSRM. In fact, GRM is selected to preserve democracy. Both BCS and BSRM can be easily derived from GSSM. For BCS, we set $R$ to an identity matrix $I_n$ and use the same size $D_i$ (or $m_i = m/c$). For BSRM, we use a fast transform matrix for $\Phi_i^B$, different sizes of $D_i$ (or $m_i \ne m_j$), and $R \in \mathbb{R}^{n \times n}$ as a random permutation of the identity matrix $I_n$.

*3.2. The RIP Condition of GSSM*

In this section, we are going to prove the asymmetric RIP condition of GSSM. From the results of RIP conditions and characteristics of BCS and BSRM, we aim for the conditions to achieve the best RIP of GSSM.

**Theorem 1.** *If c sensing matrices* $\boldsymbol{\Phi}_i = \boldsymbol{D}_i \boldsymbol{\Phi}_i^B, \forall i \leq c$ *satisfy the RIP condition*

$$(1 - \delta_i)\|\boldsymbol{z}\|_2^2 \leq \|\boldsymbol{\Phi}_i \boldsymbol{z}\|_2^2 \leq (1 + \delta_i)\|\boldsymbol{z}\|_2^2, \tag{8}$$

*for $s_*$ sparse signals $\boldsymbol{z} \in \mathbb{R}^{n_B}$, with a restricted isometry constant $\delta_i$, then GSSM matrix defined in eq. (6) satisfies the asymmetric RIP condition*

$$\frac{p}{q}(1 - \delta_*)\|\boldsymbol{\alpha}\|_2^2 \leq \left\|\frac{1}{\sqrt{q}} \boldsymbol{\Phi}_{GSSM} \boldsymbol{\alpha}\right\|_2^2 \leq (1 + \delta_*)\|\boldsymbol{\alpha}\|_2^2. \tag{9}$$

*for all $s_*$ sparse signals $\boldsymbol{\alpha} = \mathbb{R}^n$ with $s_* = \max\{s_i\} < n_B$ and $s_i$ denotes the sparsity of the sub-sampled signal $\boldsymbol{\alpha}_i = \boldsymbol{R}_i \boldsymbol{\alpha} \in \mathbb{R}^{n_B}$. In eq. (19), the value $p, q$ are the minimum and maximum number of times a sample in $\boldsymbol{\alpha}$ is selected in the set $\{\boldsymbol{\alpha}_i\}_{i=1}^c$, respectively, which are defined as*

$$p = \min \sum_{i=1}^{c} \sum_{j=1}^{n_B} \boldsymbol{R}_i(j,k) \, ; \, q = \max \sum_{i=1}^{c} \sum_{j=1}^{n_B} \boldsymbol{R}_i(j,k), \tag{10}$$
$$\forall k \in [1, 2, \cdots, n], \quad 0 < p \leq q \leq c,$$

*where $\boldsymbol{R}_i(j,k)$ denotes the j-th row and k-th columns of sub-sampling matrix $\boldsymbol{R}_i$. The extreme case of $p = 0$ is not considered as it throws away the signal information.*

**Proof.** As $\boldsymbol{\Phi}_i$ satisfies RIP for $s_*$ sparse signals, it also satisfies RIP for all $s_i \leq s_*$ with the isometric constant $\delta_i$. From eq. (6), the energy of the GSSM observation is

$$\|\boldsymbol{\Phi}_{GSSM} \boldsymbol{\alpha}\|_2^2 = \sum_{i=1}^{c} \left\|\boldsymbol{D}_i \boldsymbol{\Phi}_i^B \boldsymbol{R}_i \boldsymbol{\alpha}\right\|_2^2 = \sum_{i=1}^{c} \|\boldsymbol{\Phi}_i \boldsymbol{\alpha}_i\|_2^2. \tag{11}$$

Using eq. (11) and taking the summation of $c$ inequalities of the RIP conditions of $\boldsymbol{\Phi}_i$, we obtain

$$\sum_{i=1}^{c}(1 - \delta_i)\|\boldsymbol{\alpha}_i\|_2^2 \leq \|\boldsymbol{\Phi}_{GSSM} \boldsymbol{\alpha}\|_2^2 \leq \sum_{i=1}^{c}(1 + \delta_i)\|\boldsymbol{\alpha}_i\|_2^2. \tag{12}$$

By defining $\delta_* = \max\{\delta_i\}, i \in [1, c]$, eq. (12) becomes

$$(1 - \delta_*)\sum_{i=1}^{c}\|\boldsymbol{\alpha}_i\|_2^2 \leq \|\boldsymbol{\Phi}_{GSSM} \boldsymbol{\alpha}\|_2^2 \leq (1 + \delta_*)\sum_{i=1}^{c}\|\boldsymbol{\alpha}_i\|_2^2. \tag{13}$$

In addition, each sub-sampled signal $\boldsymbol{\alpha}_i$ consists of $n_B$ randomly selected samples from $n$ samples of $\boldsymbol{\alpha}$. Thus, for $c$ random sub-sampling, one sample can be selected multiple times. From eq. (10) and $[\boldsymbol{\alpha}(k)]^2 \geq 0$, we have

$$p[\boldsymbol{\alpha}(k)]^2 \leq \sum_{i=1}^{c}\sum_{j=1}^{n_B} \boldsymbol{R}_i(j,k)[\boldsymbol{\alpha}(k)]^2 \leq q[\boldsymbol{\alpha}(k)]^2, \tag{14}$$

where $\boldsymbol{\alpha}(k)$ represent the $k$-th element of $\boldsymbol{\alpha}$. The summation for all $n$ location from eq. (14) can be taken as

$$\sum_{k=1}^{n}[\boldsymbol{\alpha}(k)]^2 \leq \sum_{k=1}^{n}\sum_{i=1}^{c}\sum_{j=1}^{n_B} \boldsymbol{R}_i(j,k)[\boldsymbol{\alpha}(k)]^2 \leq q\sum_{k=1}^{n}[\boldsymbol{\alpha}(k)]^2. \tag{15}$$

Alternatively, the energy of sub-sampled signal $\boldsymbol{\alpha}_i = \boldsymbol{R}_i \boldsymbol{\alpha}$ is

$$\|\boldsymbol{\alpha}_i\|_2^2 = \sum_{k=1}^{n}\sum_{j=1}^{n_B}[\boldsymbol{R}_i(j,k)\boldsymbol{\alpha}(k)]^2 = \sum_{k=1}^{n}\sum_{j=1}^{n_B}\boldsymbol{R}_i(j,k)[\boldsymbol{\alpha}(k)]^2. \tag{16}$$

This is because $\boldsymbol{R}_i$ is $n_B$ randomly selected samples from the identity matrix $\boldsymbol{I}_n$, and each row has only one value of 1. From eq. (15) and (16), we have

$$\sum_{i=1}^{c}\|\boldsymbol{\alpha}_i\|_2^2 = \sum_{i=1}^{c}\sum_{k=1}^{n}\sum_{j=1}^{n_B}\boldsymbol{R}_i(j,k)[\boldsymbol{\alpha}(k)]^2 \tag{17}$$

As $\|\boldsymbol{\alpha}\|_2^2 = \sum_{k=1}^{n}[\boldsymbol{\alpha}(k)]^2$, eq. (14) becomes

$$p\|\boldsymbol{\alpha}\|_2^2 \leq \sum_{i=1}^{c}\|\boldsymbol{\alpha}_i\|_2^2 \leq q\|\boldsymbol{\alpha}\|_2^2. \tag{18}$$

By replacing this equality in eq. (13), we have

$$p(1-\delta_*)\|\boldsymbol{\alpha}\|_2^2 \leq \|\boldsymbol{\Phi}_{GSSM}\boldsymbol{\alpha}\|_2^2 \leq q(1+\delta_*)\|\boldsymbol{\alpha}\|_2^2. \tag{19}$$

By dividing all inequality to the non-negative value q, eq. (19) becomes eq. (9). Therefore, the sensing matrix $\frac{1}{\sqrt{q}}\boldsymbol{\Phi}_{GSSM}$ is said to satisfy the asymmetric form of RIP for the $s_*$-sparse signal $\boldsymbol{\alpha} \in \mathbb{R}^n$ that satisfies $s_* \leq n_B$ with the restricted isotropy $\delta_*$ and $0 < p \leq q$. GSSM satisfies the traditional RIP at $p = q$.

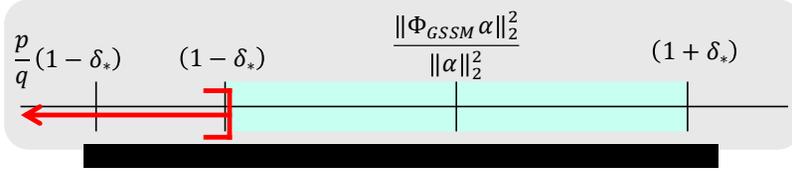

Fig. 2. Illustration of the RIP range of GSSM matrix. The smaller RIP range, the better RIP condition.

### 3.3. Analysis of the RIP Condition of GSSM

From the RIP condition in Section 3.2, GSSM does not guarantee the asymmetric RIP condition for all $s$-sparse signals in $n$ dimension ($s \ll n$) but for all $s_*$-sparse signal with $s_* = \max\{s_i\} < n_B$. This condition can be satisfied for (i) a very sparse signals with $\max\{s_i\} < s \ll n_B$, (ii) or distributed/structured sparse signals so that we can design the random selection $R_i$ satisfy $\max\{s_i\} \approx s/c < n_B$. While it is straight forward for the first case, the second case is held for structure sparse images (i.e. block sparse, structure sparse in DCT, DWT).

In this section, we aim to design an instance of GSSM (i.e., design $\boldsymbol{D}, \boldsymbol{\Phi}, \boldsymbol{R}$) to achieve better RIP condition. In order to do so, the targeted GSSM matrix should satisfy the following conditions:

**S1 – Small RIP range**. From Eq. (19), at a given $\delta_*$, GSSM achieves the smallest RIP range at $p = q$ as it is visualized in Fig. 2, and also satisfy the conventional RIP condition. This means, each sample of signal $\boldsymbol{\alpha}$ is equally and selected $cn_B/n$ number of times. We can observe that both BCS and BSRM satisfy this condition at the special case of $p = q = 1$. However, our GSSM does not limit to this special case. It should be noted that, from the definition of GSSM, we do not consider the case of $p = 0$ as it throws away signal information.

**S2 - Small $\delta_*$.** To achieve a small value of $\delta_* = \max \delta_i$, each $\boldsymbol{\Phi}_i$ should satisfy RIP with a small restricted isometric $\delta_i$. In addition, $\delta_i$ depends on the sparsity of signal $s_* = \max s_i$ ($s_i$ is the sparsity of signal $\boldsymbol{\alpha}_i$) as well as the size of sampling matrix $\boldsymbol{\Phi}_i = \boldsymbol{D}_i \boldsymbol{\Phi}_i^B \in \mathbb{R}^{m_i \times n_B}$. In general, a smaller sparsity level $s_i$ requires less measurement (i.e., smaller $m_i$) for perfect reconstruction. However, the sparsity level $s_i$ is unknown at the sampling. Therefore, the best solution to achieve S2 is

- **(S2A)** Equal size of $\boldsymbol{\Phi}_i$ (by setting $m_i = m/c$) and large block size $n_B$.
- **(S2B)** Equal sparsity level $s_i$ (by designing the sub-sampling set $\{\boldsymbol{R}_i\}_{i=1}^c$).

## 4. Restricted Structural Random Matrix

Regarding these RIP criteria, BCS satisfies S1 and S2A and BSRM satisfies S1 only. BCS uses the same block size of the sampling matrix $\boldsymbol{\Phi}_i = \boldsymbol{D}_i \boldsymbol{\Phi}_i^B \in \mathbb{R}^{m_i \times n_B}$, where $m_i = \frac{m}{c}$ $\forall i$, thus satisfies S2A. BSRM captures $c$ block signals of size $n_B = n/c$ by a sampling matrix $\boldsymbol{\Phi}_i \in \mathbb{R}^{m_i \times n_B}$; these are likely to have different sizes $m_i \neq m_j, \forall i \neq j$ due to their random generation. As a result, BSRM does not satisfy S2A. Selecting samples from the non-overlapping window in BCS or fully random samples in BSRM might meet the S2B condition for some types of signals but not for all. For random sparse signals, how to perform sub-sampling does not matter much due to the randomness. However, for compressible or structure random signal with the smoothness prior, there is possible to design sub-sampling set $\{\boldsymbol{R}_i\}_{i=1}^c$ to achieve an equally sparse level $s_i$.

Thus, we unify the advantages of both BCS and BSRM as
- S1: design $\boldsymbol{R}$ to satisfy $p = q$ (like BCS and BSRM) to have better RIP and satisfy the traditional RIP.
- S2A: use the same size $m_i$ for each block (like BCS)
- To preserve security, $\boldsymbol{R}$ is randomly generated (like BSRM)

However, further development is required since neither BCS nor BSRM satisfy the S2B condition.

Algorithm 1. Restricted Random Permutation (RRP)

| | |
|---|---|
| Input | Image size $n$, block size $n_B$, method: concat./diag. |
| Initial | $c = n/n_B$, $\boldsymbol{I} \in \mathbb{B}^{n \times n}$ is an identity matrix |
| Main | For $i = 1$ to $n_B$ |
| | $\quad \boldsymbol{t}_i = \text{randperm}(c,c) + (i-1)c$ |
| | $\boldsymbol{v} = [\boldsymbol{t}_1, \dots, \boldsymbol{t}_{n_B}]$ |
| | $\boldsymbol{V} = \text{reshape}(\boldsymbol{v}, [n_B, c])$, |
| | For $i = 1$ to $c$ |
| | $\quad \boldsymbol{U}_i = \boldsymbol{I}(\boldsymbol{V}(i,:),:)$ |
| | If method is concat $\quad$ -- for matrix $\boldsymbol{R}$ |
| | $\quad \boldsymbol{U} = [\boldsymbol{U}_1; \dots; \boldsymbol{U}_c]$ |
| | Else $\quad$ -- for matrix $\boldsymbol{D}$ |
| | $\quad \boldsymbol{U} = \boldsymbol{blkDiag}(\boldsymbol{U}_1; \dots; \boldsymbol{U}_c)$ |
| Output | Matrix $\boldsymbol{R}$ |

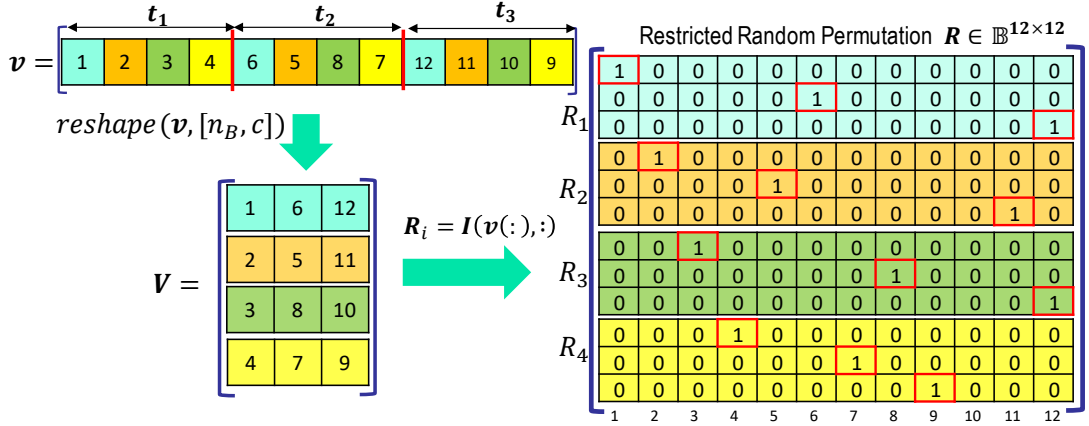

Fig. 3. Illustration of Restricted Random Permutation in Algorithm 1 with signal dimension of $n = 12$, block size $n_B = 3$, $p = q = 1$.

## 4.1. Design Matrix $R$

### 4.1.1. Restricted Random Permutation with the Global Smoothness Prior

To satisfy S2B, we designed the matrix $R$ so that the sub-sampled signals have a small and equally sparsity level. In order to do so, the smoothness prior, existed in many signals like compressible, structure sparse, etc., was utilized. For instance, in block-sparse signal, if a location has a non-zero value, then its nearby locations most likely to have non-zero values. In compressible signals, nearby samples are also highly correlated. In this work, we, however, utilized the smoothness prior in a global manner. That is, LR instances of compressible, structure sparse signals are highly correlated, equally in energy, and similar sparsity level.

Therefore, we restricted the sub-sampling matrix $R_i$ to output sub-sampled signal $x_i = R_i x$ as a random LR instance of the input signal $x$ to enforce similarity in structure among $x_i$. For the special case of $p = q = 1$, we set $c = n/n_B$ and design $c$ operators $R_i$ as shown in Fig. 3 and Algorithm 1. Notation *randperm()* and *reshape()* are functions for random permutation and reshaping, respectively. We referred this sub-sampling method as the restricted random permutation (RRP) as the input signal is permutated to create random LR instances. Our RRP matrix can be represented in terms of a binary matrix $R \in \mathbb{B}^{n \times n}$ or a random integer vector of $v \in \mathbb{N}^n$. To show the advantage of RRP over the block-by-block selection in BCS, and random permutation in BSRM, the sub-sampled instances of various signal types are visualized in Fig. 4.

For random sparse signals, there is no difference between sub-sampling methods due to the randomness of the non-zero value locations. Block-by-block sub-sampling in BCS can exploit the sparsity in the compressible signals but the sparsity level in each sub-sampled signal is likely different. Additionally, BCS is inefficient for the block-sparse signals as we have three very sparse and a very dense sub-sampled signal, thus poor S2B condition. Random permutation (RP) in BSRM can handle block-sparse signals but completely destroy the structure of compressible signal (thus not satisfy S2B for compressible signal). Our restricted random permutation preserves the structure of both compressible and block-sparse signals. Thanks to the global smoothness prior, the sub-sampled signals of RRP are highly correlated, thereby, result in a similar sparsity level for block-sparse and compressible signals in the transform domain.

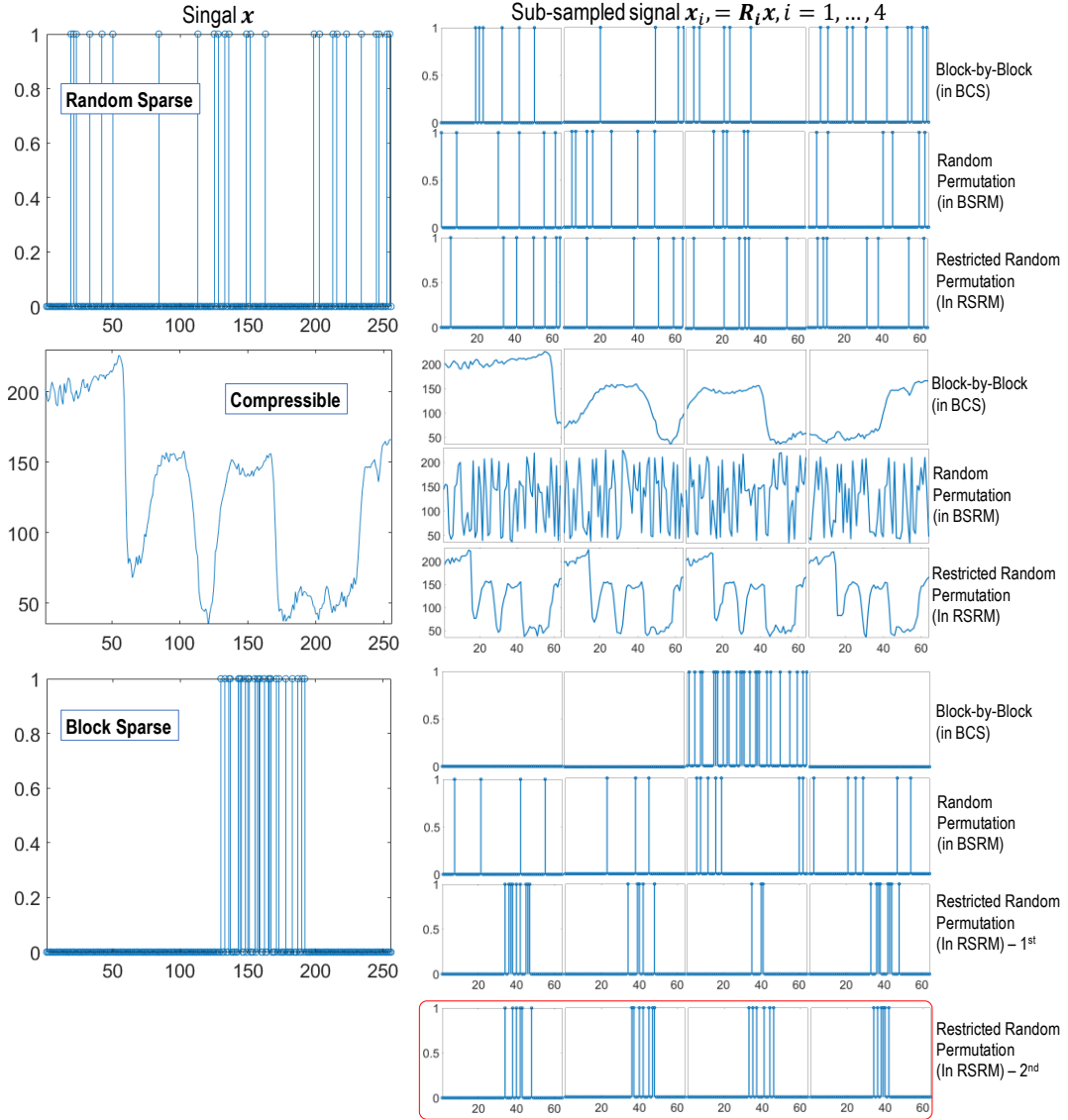

Fig. 4. Sub-sampled signals of block-by-block (in BCS), random permutation (in BSRM) and restricted random permutation (in RSRM).

### 4.1.2. Multiple Restricted Random Permutation, $c > n/n_B$ or $p=q>1$

It is well-known that the larger block-size $n_B$ is, the sparser signal is, the better RIP condition and reconstruction performance can be achieved [14, 15]. Both BCS and BSRM, however, limit $n_B$ to the relationship $n_B = n/c$ or $p = q = 1$. Therefore, the large $n_B$ can benefit RIP condition of each sampling matrix $\Phi_i$, but also result in a small number of sub-sampled signals $c$ which leads to a higher chance to waste measurements. As the to-be-sampled signal is unknown, we capture each sub-sampled signal $x_i = R_i x$ with the same number of measurements $m_i = m/c$. In the worst case of block-sparse signal in Fig. 4, BCS could waste $3m/c$ measurements to captures zero-level sparse signal, while not spending enough measurement to capture the dense signal.

Random permutation in BSRM and restricted random permutation seems to work well for all types of signals. However, there is a higher possibility of measurements waste at the large $n_B$ (or small $c$). That is, unluckily, we can partition sub-sampled signals to have very sparse or very dense. Thus, one-time division to multiple sub-sampled signal one time (i.e. $p = q = 1$) might not be enough to guarantee the S2B. In fact, the RIP condition only requires $p = q$ not the special case of $p = q = 1$ (as in BCS and BSRM). By using $p = q > 1$, we sub-sampling the whole signal multiple times and avoid the worst scenario of measurement waste for all types of signals. As visualized in Fig. 5, for block-sparse signal, the first RRP results in a sub-sampled signal that much sparser than the others. The second RRP, however, selects a different set of sub-sampled signals thus avoid wasting measurements.

This is very important because we do not assume the structure of signal in advanced. Multiple RRPs are used to have large block size $n_B$ for better performance and large $c$ to prevent the worst case of wasting measurement.

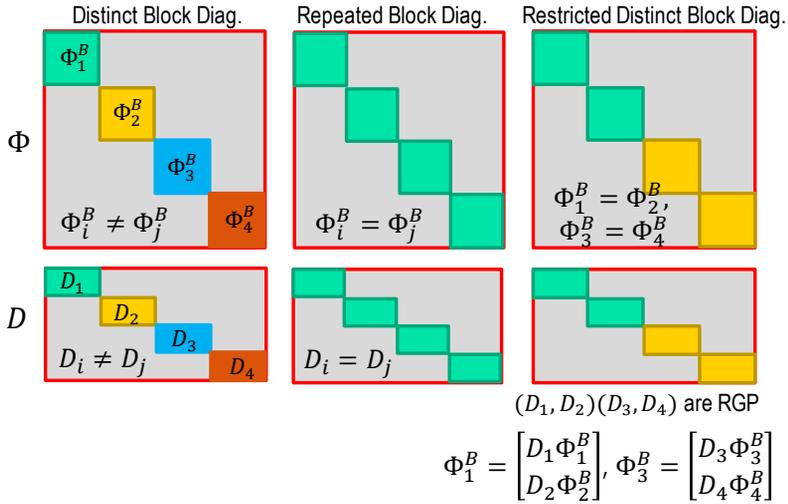

Fig. 5. Comparison of various types of block matrices $\boldsymbol{\Phi} = \boldsymbol{D}\boldsymbol{\Phi}^B$. For $\boldsymbol{\Phi}_i^B$, same colour means same matrix and vice versa. In case of $\boldsymbol{D}_i$, a matrix concatenated from all $\boldsymbol{D}_i$ of same color is equivalent to a permutation of an Identity matrix.

### 4.2. Design Matrix $\boldsymbol{\Phi}$

Since $\boldsymbol{\Phi}$ is a block diagonal matrix, its local operator $\boldsymbol{\Phi}_i$ can be difference in the *Distinct Block Diagonal* (DBD) or identical in the *Repeated Block Diagonal* (RBD) [6]. We can represent these matrices as a combination of $\boldsymbol{\Phi}_i^B$ and $\boldsymbol{D}_i$. DBD is achieved by using different sets of random matrices $\boldsymbol{\Phi}_i^B$ and sub-sampling $\boldsymbol{D}_i$. RBD is achieved by using the same $\boldsymbol{\Phi}_i^B$ and $\boldsymbol{D}_i$ for all $i$. As reported in [6], DBD shows a better RIP condition than RBD as well as better reconstruction performance at a cost of higher complexity and storage of multiple random matrices $\boldsymbol{\Phi}_B$. Additionally, $\boldsymbol{\Phi}^B$ is designed as orthogonal to save the complexity of matrix multiplication and captures signal more efficiently.

In this work, we proposed a new type diagonal matrix of *Restricted Distinct Block Diagonal* matrix (RDBD) which is a hybrid combination of DBD and RBD. Our RDBD uses each matrix $\boldsymbol{\Phi}_i^B$ multiple times. It requires significant less distinct block $\boldsymbol{\Phi}_B$ than DBD to reduce complexity. As example in Fig. 5, RBD requires one matrix $\boldsymbol{\Phi}_B$, DBD demands four matrices, but our RDBD needs only two matrices. To maintaining high quality, we should design the random selection matrix $\boldsymbol{D}$ accordingly.

### 4.3. Design Matrix $D$

Utilize all orthogonal random vector of $\Phi_i^B$ can benefit the measurement to capture different signal information. Therefore, we designed $D_i$ as RRP matrix but with the diagonal version of Algorithm 1 as shown in Fig. 5. Each set of $n/n_B$ $D_i$ corresponds to a distinct orthogonal random matrix $\Phi_i^B$. Thus, we can reduce the number of distinct random matrix $\Phi_i^B$ in comparison with the RDBD framework. Additionally, at a special case of very similar sub-sampled signals $x_i \approx x_j, i \neq j$, our method corresponds to capture a low-resolution image $x_i$ at much higher subrate. This is equivalent to favoring the low-frequency prior in [24] where authors tried to capture low-resolution images at higher subrate. However, our proposed method can favor low-frequency prior without sacrificing democracy and high-frequency contents.

### 4.4. Proposed Restricted Structural Random Matrix (RSRM)

This section summarizes our proposed method, which is named to as the restricted structural random matrix (RSRM).

**Remark 1**. *RSRM is a GSSM matrix where $c = bn/n_B$ with multiple restricted sub-sampling matrices $D_i \in \mathbb{R}^{\frac{m}{c} \times n_B}$ and $R_i \in \mathbb{R}^{n_B \times n}$ and RIP satisfied matrices $\Phi_i^B \in \mathbb{R}^{n_B \times n_B}$. $\{R_i\}_i^c$ and $\{D_i\}_i^c$ are described in Section IV*.

We can interpret RSRM as a two-step sampling process, as described in Fig. 1: (i) generate multiple random low-resolution signals and then (ii) sample each with the same number of measurements. The algorithm is given in Algorithm 2. RSRM benefits form the large block size $n_B$ and large number of sub-sampling $c$ to prevent the measurement waste. In addition, RSRM has lower complexity than fully random GRM, slightly more complexity than BCS, slightly less than BSRM with GRM structure. High reconstruction performance is also reported in the experimental results. Meanwhile, the restricted random permutation preserves the democracy property. We visualized the RSRM at the special case of $p = q = 1$ (RSRM1) in Fig. 6. We can observe four distinct brands corresponding to four sub-sampling signals. RSRM1 and BSRM sampling matrices appear more random than BCS.

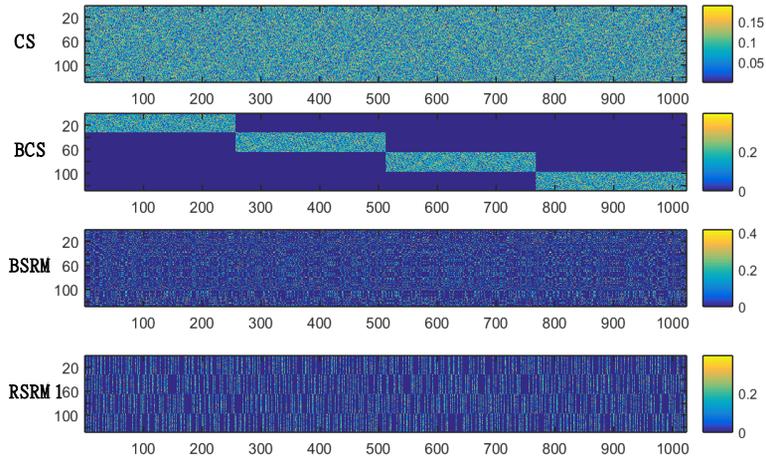

Fig. 6. Visualization of various sampling matrices ($\Phi_{CS}, \Phi_{BCS}, \Phi_{BSRM}, \Phi_{RSRM1}$) at signal dimension $n = 1024$ and block size $n_B = 256$. The randomness is observed in CS, BSRM, and RSRM. BCS is very sparse outside the diagonal, thereby, later reveals the signal energy.

Algorithm 2. Proposed Restricted Structural Random Matrix (RSRM)

| | |
|---|---|
| Input | Image size $n$, block size $n_B$, subrate $r$, and $p = q = b$ |
| Initial | No. measurements: $m = \lfloor r \cdot n \rfloor$,<br>No. sub-sampled signals: $c = b \cdot n/n_B$,<br>No. measurements per sub-sampled signal: $m_i = \lfloor m/c \rfloor^*$<br>No. GRM random matric to generate: $d = \lceil m/n_B \rceil$<br>Other: $k = 0$, and $m_{rem} = m - d * n_B$ |
| Main | **for** $i = 1:b$                                       // Generate matrix $R$<br>    $R_i$ = RRP_Gen($n, n_B, concat.$)<br>$R = [R_1; R_2; ... ; R_b]$<br>**for** $i = 1:d$                                      // Generate matrix $D$<br>    $D_i$ = RRP_Gen($n_B, m_i, diag.$)<br>$D$ = blkDiag($D_1, D_2, ..., D_d(1:m_{rem})$)<br>**for** $i = 1:c$                                      // Generate matrix $\Phi$<br>    $k = k + m_i$<br>    **if** $k >= n_B$<br>        $\Phi_B$ = GRM_Gen($n_B$)<br>    $\Phi_B^i = \Phi_B$<br>$\Phi$ = blkDiag($\Phi_B^1, \Phi_B^2, ..., \Phi_B^c$) |
| Output | RSRM matrix $D \cdot \Phi \cdot R$ |

*Remaining measurement $(m - m_i \cdot c)$ are distributed equally and randomly.
RRP_Gen(.), GRM_Gen(.) denotes function to generate RRP and GRM matrices, respectively

*4.5. RSRM with Kronecker Compressive Sensing for High Dimensional Signal and Reconstruction*

At large block size $n_B$, RSRM might still demand high complexity and storage, especially for high dimensional signals like images and videos. Thus, we incorporated RSRM to separable sampling. We used two RSRM matrices for vertical and horizontal sampling. We used group sparse representation (GSR) [41] to represent image as

$$x = \Psi_G \circ \alpha_G, \quad (20)$$

where $\Psi_G, \alpha_G$ denote the group dictionary and group sparse coefficients, respectively. Utilizing the group sparsity as the cost function $J = \|\alpha_G\|_0$, the restoration of KCS is given as

$$\arg\min_{X, \alpha_G} = \|\Phi_L X \Phi_R^T - Y\|_F^2 + \lambda \|\alpha_G\|_0 \quad s.t. \quad \Psi_G \circ \alpha_G, \quad (21)$$

where $X$ is a 2D representation of 1D signal $x$ via a vectorized function as $x = vect(X)$ and $\|\cdot\|_F$ denotes the Frobenious norm. Similar to [9], we use split Bregman optimization to divide and solve Eq. (21).

Table 2. Comparison between CS, BCS, SRM, and RSRM at image size $n$ and block size $n_B$

| Param | CS | BCS | BSRM | RSRM |
|---|---|---|---|---|
| Block size | $n_B = n$ | $n_B < n$ | $n_B < n$ | $n_B < n$ |
| $p = q$ | $b = 1$ | $b = 1$ | $b = 1$ | $b \geq 1$ |
| No. block | $c = 1$ | $c = n/n_B$ | $c = n/n_B$ | $c = bn_B/n$ |
| No. meas. | $m_i = m$ | $m_i = m/c$ | $m_i \neq m_j$ | $m_i = m/c$ |
| $R$ | $I_n \in \mathbb{B}^{n \times n}$ | $I_n \in \mathbb{B}^{n \times n}$ | Rand. Perm., $\in \mathbb{B}^{n \times n}$ | RGP, $\in \mathbb{B}^{cn_B \times n}$ |
| $\Phi_B$ | $\in \mathbb{R}^{n \times n}$<br>1 GRM | $\in \mathbb{R}^{n_B \times n_B}$<br>$c$ block of GRMs | $\in \mathbb{R}^{n_B \times n_B}$<br>$c$ block of GRMs | $\in \mathbb{R}^{n_B \times n_B}$<br>$\lceil m/n_B \rceil$ block of GRM |
| $D$ | $\in \mathbb{B}^{m \times n}$ | $\in \mathbb{B}^{m \times n}$ | $\in \mathbb{B}^{m \times n}$ | RGP, $\in \mathbb{B}^{m \times cn_B}$ |

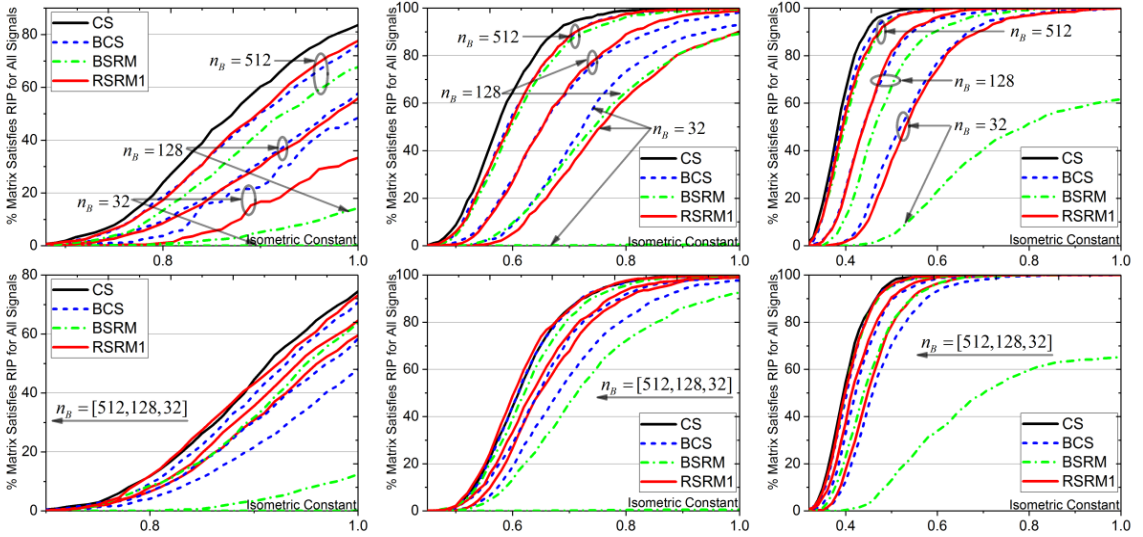

Fig. 7. Percentages of matrices of BCS, BSRM, and RSRM1 that satisfy RIP for all random sparse signals of size $n = 1024$ versus isometric constant $\delta$. The sparsity level is s = 16 and 32 in the first and second rows, respectively. The number of measurements is $m = 32, 64,$ and $128$ in the first, second, and third columns, respectively. The curve towards the top left shows the better RIP condition.

## 5. Experimental Results

### 5.1. Evaluation of the RIP Condition

This section evaluates the RIP conditions of CS, BCS, BSRM, and RSRM at $p = q = 1$ (RSRM1). All sampling matrices are derived from GSSM with the configuration in Table 2 and GRM is used as the sampling matrix $\Phi_B$. Similar to [6], we count the number of sensing matrices that satisfy the RIP condition at a given $\delta \in (0, 1)$ for all 1000 signals $x$ of size $n = 1024$. The signals are random sparse, block-structure sparse [39], and compressible signals. The results are shown in Fig. 7, 8, and 9. Each subfigure contains 10 plots: one for CS (black) and three plots each for BCS (blue dash), BSRM (green dotted dash), and RSRM1 (red solid line) at different block sizes of $n_B = 128, 256,$ and $512$. We use the location of the plot to distinguish between $n_B$ plots; plots that are closer to the top left have a larger block size $n_B$. In general, the greater the number of measurements $m$, the better the RIP condition.

### 5.1.1. Random Sparse Signals

We generated 1000 sparse signals with the sparse level of $s = 16, 32,$ and $64$ by assigning a non-zero value for $s$ random locations. Since the non-zero is random, the difference in the sparsity level between $n_B$ sub-samples becomes less at larger $s$. Therefore, the sub-sampling method plays little role. All methods BCS, BSRM, and RSRM1 are likely to satisfy S2B for relatively less sparse signals (i.e., large $s = 32$) as their RIP gaps are shorted in comparison with RIP condition of fully random GRM (CS). Satisfying S2A condition, BCS and RSRM1 were expected to show better RIP condition than BSRM. At $s = 16$ and $n_B = 32$, the order of decreasing RIP conditions is CS, BCS, RSRM1, and BSRM. In fact, RSRM1 shows almost identical performance to BCS at $n_B = 128, 512$ and/or $m = 64, 128$. However, at $s = 32$, RSRM1 even shows a similar RIP condition to fully random CS. In general, RSRM1 has better RIP condition than both BCS and BSRM for all block sizes and measurements.

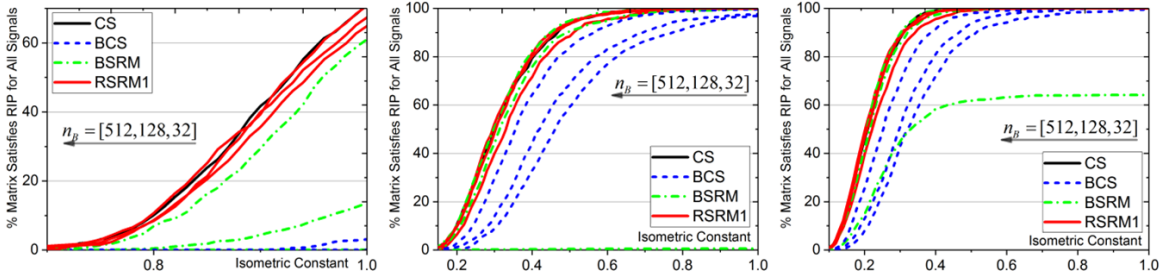

Fig. 8 Percentages of matrices (BCS, BSRM, and RSRM1) that satisfy the RIP condition (for all 1000 block-sparse signals of size $n = 1024$, block-sparse size = 256, block sparsity = 2) versus the isometric constant $\delta$ at various measurements (left to right $m = 32, 64,$ and 128) and various block sizes $n_B$.

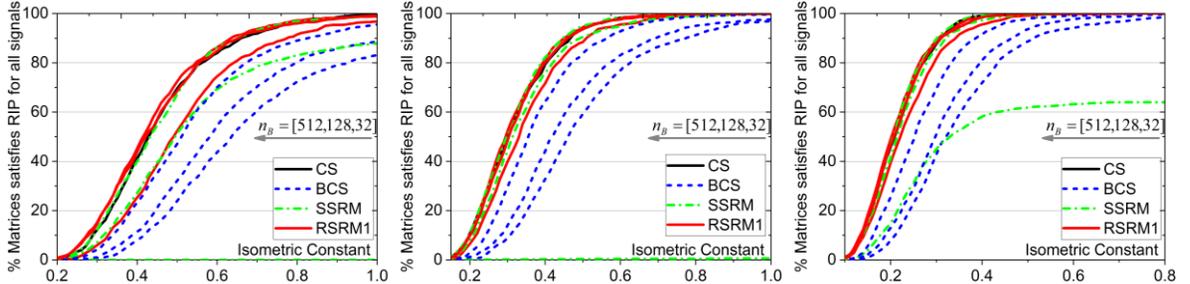

Fig. 9 Percentages of matrices (BCS, BSRM, and RSRM1) that satisfy the RIP condition for all compressible signals of size $n = 1024$, block-sparse size = 256, and block sparsity = 2, versus the isometric constant $\delta$ at various measurements (left to right $m = 32, 64,$ and 128) and block size $n_B$.

*5.1.2. Block-Sparse Signals*

In block-sparse signals, non-zero values are located in a few non-overlapping blocks [39]. This section generates block-sparse signals at sparsity levels $s = 16, 32,$ and 64 with a block-sparse size of 256 and a block sparsity level of 2. That is, we generate $s$ non-zero values located in two random blocks of size 256 and left the other two blocks all zero. BCS fails to satisfy S2B as one of its sub-sampled signals has zero sparsity level, $s = 0$. With a random permutation, the sub-sampled signals in BSRM are less likely to has $s = 0$ than BCS. Our RSRM1's sub-sampled signals are not only block sparse but also likely to have similar sparsity level, thus offers better RIP, especially at small measurement $m$ and small block size $n_B$. This argument also agreed with the conclusion for BCS in [6].

*5.1.3. Compressible Signals*

In practice, compressible signals (e.g., audio, images, and video) are very common. Additionally, we can always approximate a compressible signal $x$ with an $s$-sparse signal $\alpha$ in a given sparsifying domain [40], like DCT and DFT. This section generated 1D compressive signals by extracting random rows and columns of test sequences in conventional video coding standards: Cactus, Kimono, and QBTerrace. Compressible signals have strong smoothness prior so that the block signals in BCS are also compressible and can be approximated with the sparsity level $s_i$. However, the sparsity level between these block signals is likely different. The adaptive BCS scheme was proposed to take advantage of the different sparsity levels [38]. Unfortunately, the sparsity level is unknown in the sampling stage. Alternatively, BSRM destroys the smoothness prior of the signal due to the random selection, especially at small $n_B$. Meanwhile, RSRM1 preserves signal structures thanks to the smoothness prior and produces a better RIP condition than BCS and BSRM, especially at the small block size $n_B$ and measurement rate $m$.

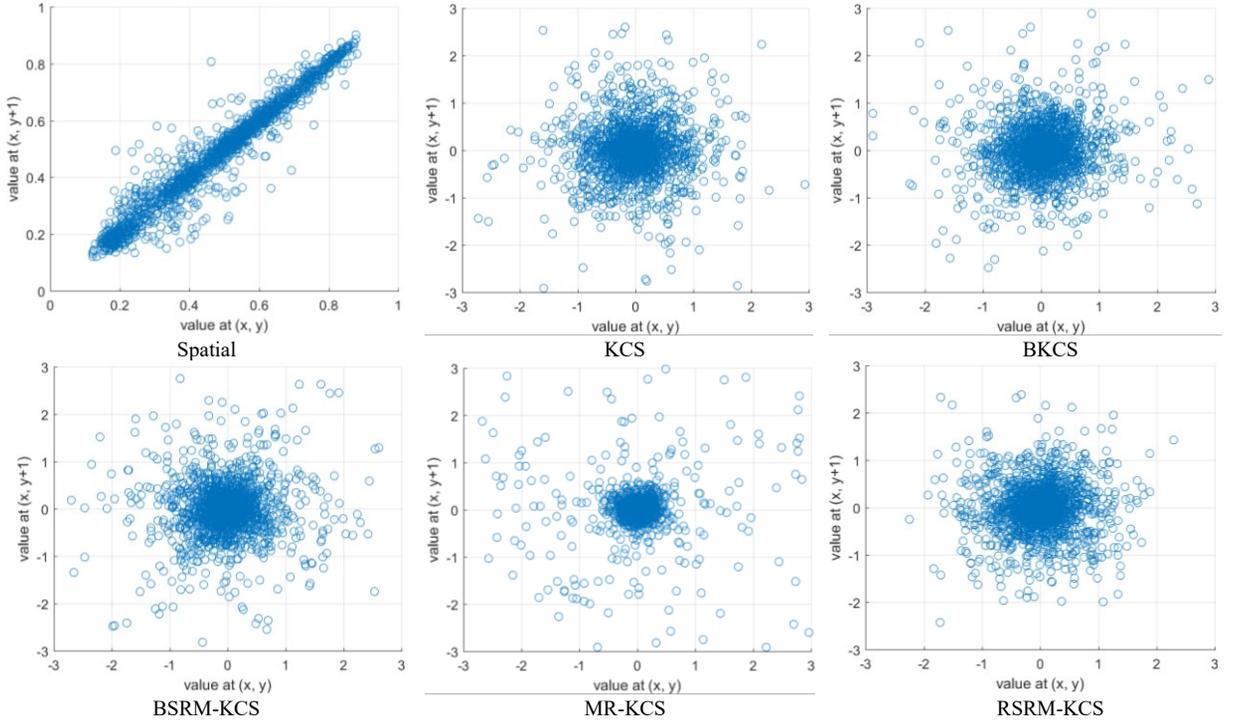

Fig. 12 Correlation between to nearby pixels/measurements Lena at subrate 0.1 in spatial and compressed domains.

## 5.2. Evaluation of the Security/Democracy

### 5.2.1. Evaluate Secured Measurements

With the restricted random permutation, each sub-sampled signal in RSRM is an LR version of the original signal with equal energy. Also, the structure of signals is also preserved in the LR images. Additionally, each block is sampled with a GRM matrix. Therefore, there is no measurement loss and measurements of RSRM are also equally important. Fig. 12 demonstrates the correlation between nearby measurements. There is a strong correlation between the nearby pixels in the spatial domain but not the measurement domains. Also utilizing signal prior, but RSRM's observations are less correlated than the multi-scale sampling with MR-KCS. In Fig. 13, while low-frequency are easily located in MRKCS, no visible difference is shown in the other sampling scheme. Additionally, by evaluating the energy of each block measurement, an LR image is revealed in the BCS scheme.

Table 3. Storage requirement of various matrices for image of size 256×256, subrate 0.25, $n_B = 128$

|  | Frame-based sensing | | | | | Separable (Kronecker) sensing | | | | |
|---|---|---|---|---|---|---|---|---|---|---|
|  | CS | BCS | BSRM | RSRM1 | RSRM4 | KCS | BKCS | BSRM | RSRM1 | RSRM4 |
| **R** | 0 | 0 | $256^2$ | $256^2$ | $4 \times (256^2)$ | 0 | 0 | $2 \times (256)$ | $2 \times (256)$ | $8 \times (256)$ |
| **D** | 0 | 0 | $128^2$ | $128^2$ | $128^2$ | 0 | 0 | $2 \times (128)$ | $2 \times (128)$ | $2 \times (128)$ |
| **Φ** | $128^2 \times 256^2$ | $0.25 \times 256^2$ | $0.5 \times 256^2$ | $0.25 \times 256^2$ | $0.25 \times 256^2$ | 2(128×256) | 2(128×128) | 2(128×128) | 2(128×128) | 2(128×128) |

Values represent number of *integer elements* in **R** and **D** (in vector format) and number of floats elements in **Φ**

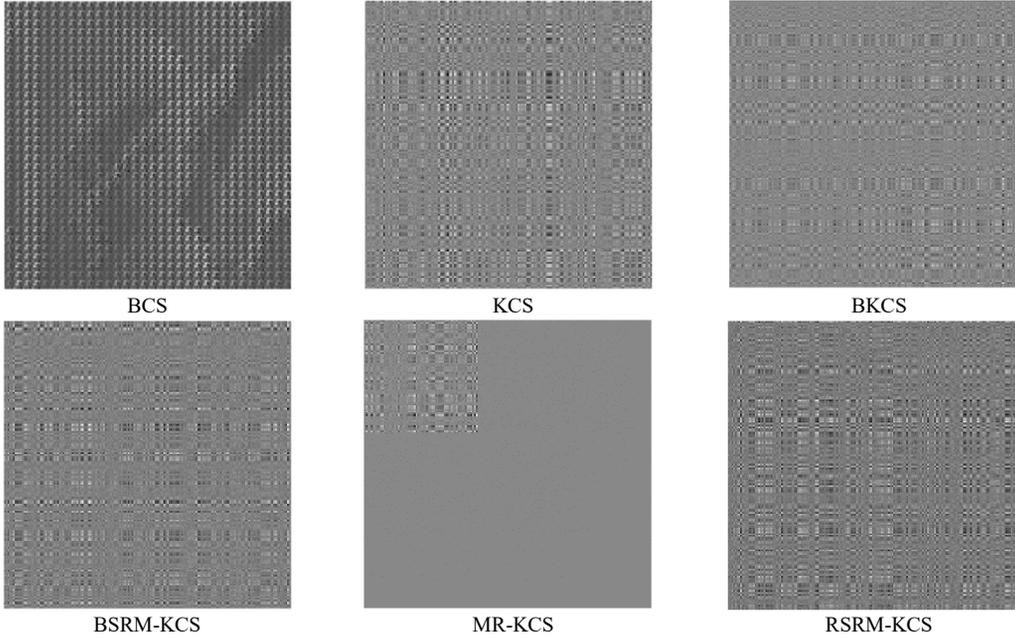

Fig. 13 Measurements of various sampling schemes at subrate 0.1. BCS measurement are captured at block size 8×8, 100 (over 102) measurements at each block is reshaped to 10×10 measurement block and concatenated to form the larger image.

### 5.2.2. Complexity of Sampling Matrices

The complexity of sampling matrices is evaluated in terms of storage requirements. As in Table 3, the float parameter represents the size of the random matrix and binary parameter represent the size of the permutation matrices $D, R$. At subrate $r = 0.25$, RSRM required only $0.25 \times 256^2$ float parameters, similar to BCS and less than BSRM. All frameworks (BCS, BSRM, RSRM) used the same size of matrix $D$ but, RSRM demanded little storage for matrix $R$. The larger number $p = q = b$ is, the more storage is required. However, the increment storage for $R$ (in binary) is much smaller than that of the matrix $\Phi$ (in floats) and still significant less than the fully random CS. On the other hand, the separable sampling scheme greatly reduces the complexity of the conventional frame-based sensing. Also, RSRM only slightly increases complexity for the binary matrix $R$ in comparison with BCS and BSRM.

We also evaluate the complexity of the sensing operation and reconstruction in terms of the number of additions and multiplication for the sampling operation $y = \Phi x$. The conventional frame-based CS requires $m \times n^2$ additions and multiplications to captures $m$ measurements. As $D$ and $R$ are determined binary matrices, they can be implemented without matrix multiplication (i.e., pre-select/rearrange sampling matrices and signals). Therefore, all other matrices BCS, BSRM, and our RSRM requires $m \times n_B^2$ additions and multiplications for acquisition. Our RSRM at $p = q > 1$ only adds more storage but maintain the same computational complexity as BCS and BSRM, regardless the value $p = q = b$.

It should be noted that the BSRM for comparison is BSRM with diagonal GRM matrix. The conventional BSRM in the original work [12] requires fewer complexity thanks to the fast transform. It only needs to store a transform matrix of 128×128. The matrix multiplication is also faster with a fast transform. Therefore, our RSRM would have little more storage requirement with the same computation as BCS and BSRM with GRM but little more complex than BSRM with a fast transform. In exchange, RSRM can preserve democracy property.

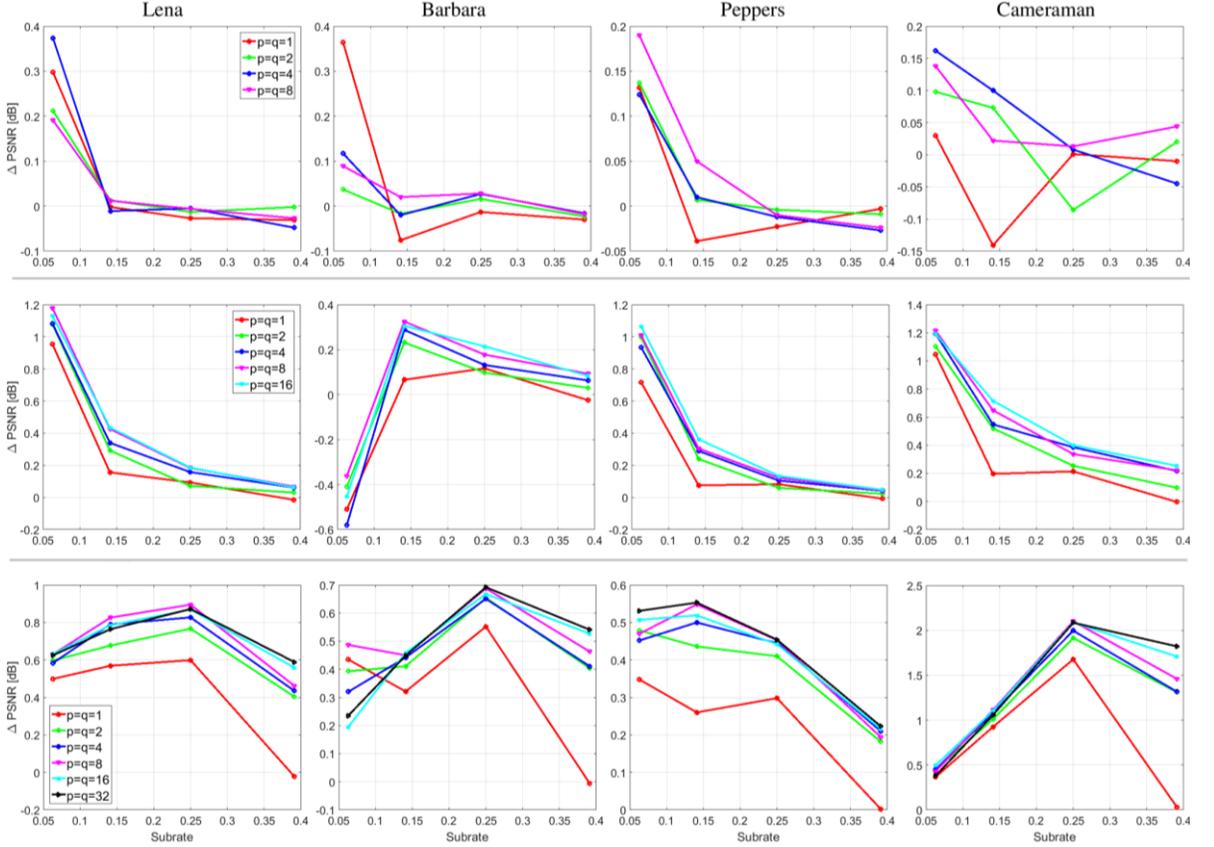

Fig. 14. ΔPSRN [dB] of RSRM-KCS over KCS at various $p = q$ and $n_B = 64, 128,$ and $256$ for the first, second, and last rows, respectively.

### 5.2.3. One-Time Sensing with Triple Level Protection

Even though CS measurements reveal the signal energy, it enables computational security at a certain level with low power complexity which is essential for resource-constrained applications like (visual) sensor network. By treating the measurement matrix as the security key, sampled signal as the plaintext, and compressed measurements as ciphertext [32-34, 50]. Sensing matrix and measurements are sent to the receiver through a private and public channel, respectively. In this scenario, attackers can perform a plain attack [33] (i.e., manipulate input and capture the corresponding output). Since CS is a linear sampling technique, attackers can guess the sensing matrix after a certain amount of time. In this regard, the best security system is one-time sampling [33] by changing and resending the sampling matrices frequently. This approach, however, causes significant overhead over the private channel. Bi-level protection [36] is proposed to reduce the transmission overhead by using both a sensing matrix and a sparsifying basis as primary keys. The sparsifying matrix is very limited to a few bases like DCT, DWT, etc. In our RSRM, we use $\boldsymbol{D}, \boldsymbol{\Phi},$ and $\boldsymbol{R}$ as primary keys to provide triple-level protection. Changing the binary matrices $\boldsymbol{R}, \boldsymbol{D}$ instead of $\boldsymbol{\Phi}$ significantly reduces the overhead over the private channel. Note that, even though $\boldsymbol{R}, \boldsymbol{D}$ is restricted to random LR sub-sampling, the number of possible spaces for $\boldsymbol{R}$ is still significantly large, especially for high dimensional signal and large block size. For instance, if $n_B = n/2$, then there is $2^{n/2}$ possible combination of $\boldsymbol{R}$.

Table 3. PSNR [dB] and SSIM reconstruction performance of various sensing schemes

|  |  | DR²Net | CSNet | BCS-MH | BCS-GSR | KCS-DETER | KCS-GSR | KCS-RSRM |
|---|---|---|---|---|---|---|---|---|
| Lena | 0.01 | 20.69/0.582 | - | 17.21/0.490 | 20.13/0.613 | 13.83/0.413 | 20.51/0.638 | **23.60/0.680** |
|  | 0.04 | 25.17/0.703 | - | 25.76/0.747 | 27.34/0.795 | 26.79/0.756 | 28.74/0.808 | **29.76/0.827** |
|  | 0.10 | 28.64/0.800 | 32.15/0.879 | 29.76/0.840 | 30.97/0.866 | 31.98/0.863 | 32.63/0.876 | **33.31/0.887** |
|  | 0.20 | - | 35.12/0.924 | 32.82/0.894 | 34.44/0.914 | 35.08/0.910 | 35.57/0.915 | **36.41/0.925** |
|  | 0.25 | 32.70/0.879 | - | 34.72/0.908 | 35.57/0.926 | 36.09/0.922 | 36.52/0.926 | **37.43/0.935** |
|  | 0.30 | - | 37.32/**0.945** | 36.24/0.919 | 36.47/0.936 | 36.90/0.932 | 37.31/0.935 | **38.00**/0.941 |
| Barbara | 0.01 | 18.65/0.439 | - | 15.30/0.351 | 17.18/0.430 | 13.19/0.293 | 17.61/0.416 | **19.28/0.453** |
|  | 0.04 | 21.14/0.558 | - | 21.03/0.568 | 21.14/0.619 | 20.62/0.507 | 22.93/0.666 | **23.17/0.643** |
|  | 0.10 | 22.52/0.647 | 24.28/0.716 | 27.74/0.839 | 30.47/0.905 | 26.70/0.785 | 31.57/0.907 | **32.04/0.915** |
|  | 0.20 | - | 25.25/0.784 | 31.49/0.912 | 34.51/0.947 | 33.00/0.924 | 34.95/0.942 | **35.70/0.949** |
|  | 0.25 | 25.20/0.779 | - | 32.54/0.926 | 35.70/0.955 | 34.33/0.938 | 36.04/0.950 | **36.88/0.958** |
|  | 0.30 | - | 26.59/0.853 | 33.67/0.937 | 36.72/0.962 | 35.42/0.948 | 36.94/**0.967** | **37.71**/0.962 |
| Peppers | 0.01 | 20.12/0.566 | - | 16.64/0.472 | 19.58/0.584 | 13.82/0.411 | 20.30/0.629 | **22.89/0.671** |
|  | 0.04 | 24.63/0.684 | - | 25.63/0.718 | 28.07/0.786 | 26.66/0.745 | 28.98/0.799 | **29.82/0.807** |
|  | 0.10 | 28.32/0.769 | 32.05/**0.857** | 30.25/0.814 | 31.75/0.843 | 32.09/0.841 | 32.10/0.842 | **32.42**/0.845 |
|  | 0.20 | - | 34.42/**0.891** | 32.83/0.860 | 34.12/0.880 | 34.38/0.878 | 34.46/0.877 | **34.89**/0.881 |
|  | 0.25 | 32.16/0.834 | - | 33.62/0.874 | 34.98/0.893 | 35.13/0.890 | 35.24/0.890 | **35.70/0.894** |
|  | 0.30 | - | 35.84/**0.911** | 34.28/0.887 | 35.64/0.905 | 35.80/0.901 | 35.95/0.902 | **36.31**/0.905 |
| Camera-man | 0.01 | 19.68/0.622 | - | 16.36/0.497 | 18.95/0.642 | 14.00/0.411 | 20.17/0.632 | **22.68/0.694** |
|  | 0.04 | 23.84/0.744 | - | 23.96/0.628 | 25.68/0.823 | 28.03/0.821 | 29.21/0.842 | **30.23/0.862** |
|  | 0.10 | 28.46/0.848 | 31.14/0.918 | 29.86/0.712 | 32.12/0.913 | 33.47/0.917 | 33.11/0.905 | **34.44/0.927** |
|  | 0.20 | - | 34.59/0.961 | 33.88/0.797 | 37.15/0.958 | 37.19/0.954 | 37.38/0.952 | **36.57/0.969** |
|  | 0.25 | 34.10/0.939 | - | 35.28/0.948 | 39.04/0.969 | 38.42/0.962 | 38.93/0.964 | **41.72/0.979** |
|  | 0.30 | - | 37.47/0.976 | 36.37/0.953 | 40.58/0.977 | 39.53/0.968 | 40.24/0.971 | **42.54/0.981** |
| Mandr-ill | 0.01 | 17.67/0.248 | - | 14.04/0.190 | 16.10/0.242 | 12.14/0.168 | 16.13/0.228 | **17.34/0.242** |
|  | 0.04 | 18.83/0.335 | - | **19.15/0.344** | 18.68/0.364 | 17.86/0.297 | 17.85/0.304 | 18.36/0.324 |
|  | 0.10 | 20.18/0.455 | **22.26/0.593** | 20.44/0.469 | 19.93/0.580 | 19.92/0.443 | 19.50/0.445 | 20.12/0.491 |
|  | 0.20 | - | **24.08/0.749** | 22.36/0.647 | 22.22/0.682 | 21.98/0.597 | 21.47/0.590 | 22.52/0.659 |
|  | 0.25 | 22.29/0.638 | - | 23.24/0.705 | 23.11/0.734 | 23.56/0.654 | 22.45/0.647 | 23.53/**0.716** |
|  | 0.30 | - | **25.72/0.833** | 23.99/0.745 | 23.92/0.775 | 23.68/0.700 | 23.39/0.696 | 24.40/0.751 |
| Goldhill | 0.01 | 21.54/0.467 | - | 15.79/0.361 | 19.94/0.464 | 14.69/0.347 | 19.93/0.464 | **21.83/0.489** |
|  | 0.04 | 24.50/0.578 | - | 24.46/0.580 | 25.26/0.630 | 23.67/0.566 | 24.30/0.587 | **25.22/0.617** |
|  | 0.10 | 27.11/0.694 | **30.43/0.810** | 27.49/0.717 | 28.28/0.752 | 27.54/0.715 | 27.17/0.703 | 27.83/0.732 |
|  | 0.20 | - | **32.70/0.883** | 30.19/0.817 | 31.40/0.848 | 30.44/0.816 | 30.53/0.815 | 31.64/0.851 |
|  | 0.25 | 30.30/0.811 | - | 31.13/0.844 | 32.50/0.875 | 31.46/0.845 | 31.76/0.850 | **32.79/0.878** |
|  | 0.30 | - | **34.48/0.919** | 31.88/0.864 | 33.35/0.893 | 32.36/0.868 | 32.77/0.874 | 33.66/0.895 |
| Boats | 0.01 | 19.69/0.458 | - | 15.86/0.361 | 18.03/0.440 | 12.97/0.308 | 18.55/0.450 | **20.26/0.475** |
|  | 0.04 | 22.55/0.561 | - | 22.43/0.580 | 23.40/0.623 | 22.77/0.575 | 23.60/0.610 | **24.24/0.632** |
|  | 0.10 | 25.64/0.688 | **29.07/0.812** | 26.12/0.717 | 27.55/0.773 | 27.24/0.738 | 27.07/0.732 | 27.82/0.762 |
|  | 0.20 | - | **32.05/0.884** | 29.33/0.817 | 31.34/0.862 | 30.74/0.837 | 30.98/0.841 | 32.00/0.865 |
|  | 0.25 | 29.47/0.811 | - | 30.32/0.844 | 32.71/0.887 | 31.96/0.864 | 32.22/0.868 | **33.28/0.888** |
|  | 0.30 | - | 33.98/**0.911** | 31.21/0.864 | 33.72/0.904 | 32.94/0.884 | 33.27/0.888 | **34.09**/0.901 |
| Man | 0.01 | 20.11/0.459 | - | 16.71/0.361 | 18.98/0.453 | 14.84/0.362 | 19.67/0.484 | **21.43/0.518** |
|  | 0.04 | 23.65/0.954 | - | 23.68/0.602 | 24.57/0.660 | 24.08/0.617 | 24.34/0.632 | **25.20/0.659** |
|  | 0.10 | 26.51/0.714 | **29.82/0.833** | 26.77/0.729 | 27.35/0.781 | 27.95/0.758 | 27.35/0.742 | 28.21/0.774 |
|  | 0.20 | - | **32.55/0.907** | 29.33/0.818 | 30.63/0.867 | 31.00/0.848 | 31.18/0.852 | 32.39/0.885 |
|  | 0.25 | 29.97/0.833 | - | 30.22/0.844 | 31.28/0.893 | 32.18/0.875 | 32.67/0.885 | **33.96/0.913** |
|  | 0.30 | - | 34.52/**0.939** | 30.98/0.864 | 32.90/0.913 | 33.21/0.895 | 33.91/0.907 | **34.94**/0.925 |
| Average |  | - | - | 26.76/0.712 | 28.45/0.770 | 27.37/0.718 | 28.74/0.755 | **29.80/0.778** |

The best results are in **bold**.

Enabling simultaneous sampling, compression, and security, CS enables very low power and simple linear encoder, thus, becomes very attractive for the Internet of Things (IoT) applications [49, 50]. Additionally, the low computation capacity of IoT devices cannot handle a very large sampling matrix, thus block-based and structure frame-based sensing is favored.

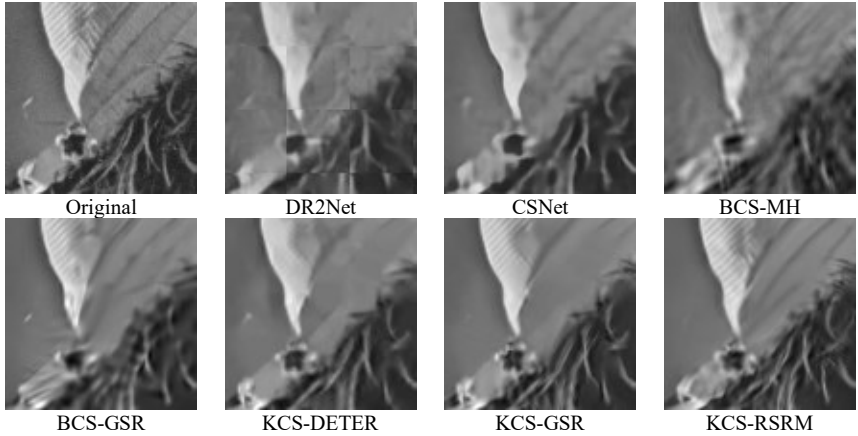

Fig. 14. Visual quality comparison of various sampling schemes for the Lena image at subrate 0.1.

However, further simplicity comes with the cost of reducing reconstructed image quality as well as compromise the security property. Fortunately, with our RSRM framework, reconstruction quality can be further improved without compromise security property at a cost of little extra complexity. Additionally, by pre-rendering multiple sets of a completed of restricted permutation matrices $\{R_i\}_{i=1}^{k_R}, \{D_i\}_{i=1}^{k_D}$, or utilizing circulant structure. Instead of sending the sampling matrix $R_i, D_i, \Phi_i^B$, we can send the indexes of matrices (already known in both encoder and decoder) to further reduce transmission overhead.

### 5.3. Reconstruction Performance of RSRM

This section presents the $\Delta$PSNR [dB] results of the proposed RSRM under Kronecker sampling (KCS-RSRM) in comparison with the conventional KCS with GSR reconstruction (KCS-GSR) shown in Fig. 8. Surprisingly, our KCS-RSRM outperforms KCS-GSR using the same GSR reconstruction, especially at low subrate.

#### 5.3.1. Block size $n_B$

Similar to 1D experiments, the performance improves as the size of $n_B$ increases. In general, RSRM shows the best gain at a large block size $n_B = 256$, with up to 2.1 dB gain in Cameraman at a subrate 0.25. However, there is an exception for a very small subrate. At a subrate of 0.05, RSRM shows a significant PSNR gain at $n_B = 128$ for smooth and complexly textured images (Lena, Peppers, and Cameraman, except Barbara which has strong local smoothness. We select $n_B = 256$ as the best solution as RSRM shows gains for all test images.

#### 5.3.2. Number of sub-sampled signals $c = bn_B/n$ and $p = q = b$

As RSRM does not limit to $p = q = b = 1$. We evaluate $b = 1, 2, 4, 8,$ and $16$. At block size $n_b$, number of sub-sampling blocks is $c = bn_B/n$. A larger value of $b$ is preferred to prevent the worst case of the measurement waste for the block-sparse and compressible signals. However, a large value $b$ leads to smaller number of measurements for each $m_i = m/c$, which subsequently affects the RIP condition of $\Phi_i^B$. Fig. 14 shows that increasing the value of $b$ leads to better performance, especially at a high subrate. However, increasing the size of $b$ also results in larger $D, \Phi,$ and $R$, which creates more complexity. We select $b = 8$ for our RSRM.

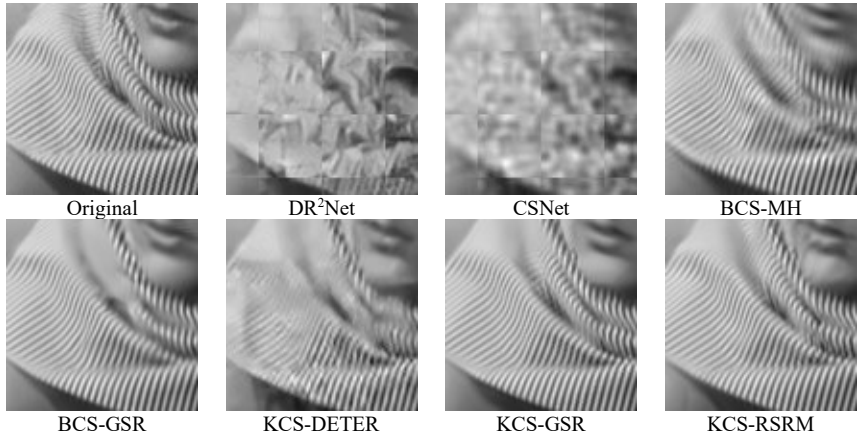

Fig. 15. Visual quality comparison of various sampling schemes for the Barbara image at subrate 0.1.

## 5.4. Comparison with State-of-the-Art Sensing Schemes

We used RSRM under the KCS with $p = q = 8$ and GSR reconstruction (named KCS-RSRM) is compared with state-of-the-art algorithms: multiple-scale BCS (MH-BCS [23]), BCS with GSR recovery (BCS-GSR [40]), KCS with GSR recovery (KCS-GSR) [11], and recent deep learning BCS frameworks with DR$^2$Net [42] and CSNet [43].

As in Table 3, KCS-RSRM produces the best PSNR performance compared with other reconstruction schemes, especially at low subrate. Surprisingly, KCS-RSRM even improves the results by 0.78 dB in average compared to KCS-GSR (which uses GRM as the sampling matrix, same reconstruction) with up to 3.09 dB gain at a low subrate of 0.01 for the Lena, and 1.08 at high subrate 0.03 for the Cameraman. For a structure image (like Barbara) or a very smooth signal (like Peppers), KCS-RSRM produces an average gain of 0.5 dB. Compared to recent deep-learning methods, KCS-RSRM offers better performance for complex images (e.g., Lena), images with a strong local structure (e.g., Barbara), highly compressible images (e.g., Cameraman), and smooth images (e.g., Peppers). As CSNet joints trained sampling and reconstruction at a corresponding subrate, its sampling matrix tends to capture more low-frequency information which is similar to exploit low-frequency prior in the multi-scale sampling scheme. With the external deep-learning prior, CSNet shows better reconstruction for the Mandrill, Boats, Goldhill, and Man images.

From Fig. 14 and Fig. 15, KCS-RSRM demonstrates the best visual quality. We also can observe that CSNet favors low-frequency images and experiences losses in high-frequency content. The reconstruction results of CSNet are blurred at the edges and in textured areas. Alternatively, KCS-RSRM is able to reconstruct sharp edges and structured areas (e.g., Lena's hair and Barbara's neck) better than KCS-GSR.

## 5.5. Discussions and Future Work

Despite showing the asymmetric RIP condition of GSSM for $p \leq q$, this work mainly evaluated the special case of $p = q$. By doing so, all samples are equally important, thus preserve democracy property. However, the current RSRM framework is single-scale sampling which can be extend to multi-scale. That is, we could consider important samples (i.e., edges, texture) more often than the others, thus results in the case of $p < q$. The multi-scale version of RSRM can further improve the reconstruction quality but at a cost of compromise democracy property. This extension is out of the scope of this paper and left for the future work.

As a generalized form BCS and BSRM, and simplified of frame-based sensing, our proposed sampling matrix RSRM could be used anywhere that those techniques are applicable. Firstly, it is true that we cannot visit the signal multiple times in practice, thereby difficult to choose the pixels to process. We design **R** as a set of restricted random permutations given the signal dimension and block size, and regardless the signal structure. Additionally, this is not the drawback of our method only but a well-known problem of CS. CS is a sequential sampling scheme, i.e. measurements are captured sequentially. Thus, implementation of the CS system for practical applications remains challenging, especially for fast-changing signals. On the other hand, the dimension of the to-be-sampled signals is assumed for the implementation. The signal resolution is defined by the resolution of the digital micromirror device (DMD) in the single-pixels camera [37], by the resolution of the coded pattern in the lensless imaging [55]. Therefore, it is straight forward to implement our RSRM to existing CS framework.

In single pixel-camera, we can pre-define locations of $c$ random LR images, then redirect the DMD at the corresponding locations to the corresponding multiple single-pixels. Each single-pixel captures measurements of a random LR image. This is for the case of $p = q = 1$ with a single RRP. For the case of multiple RRP, $p = q > 1$, multiple light splitters can be used with multiple DMDs and multiple single-pixels. Therefore, similar to BCS, RSRM enables parallel acquisitions to reduce sampling time. On the other hand, we also can implement RSRM under lensless imaging [55] system. We can sequentially capture measurements of each random LR image – specified by the coded pattern. As conventional lensless imaging is also sequential sampling, using our RSRM matrix results in the same acquisition time.

According to [50], our framework can be applied for image/video and audio security as well as outsourcing security over the cloud [51]. While many applications are considered compressive sensing for security applications only where the signal is known, our approach is considered the security during the sampling process which is closer to physical level security and/or camera side security [56]. More details analysis on the security applications as well as further advance the storage requirement with circulant structure of **R** and **D** are left for future work.

## 6. Conclusion

In this work, we proposed a novel restricted structurally random matrix (RSRM) that combines the advantages of block-based and structured frame-based frameworks and utilizes the global smoothness prior. Our RSRM offers a better RIP conditions than BCS and BSRM for various types of signals, including random sparse, block-sparse, and compressible. Especially for compressible signals, RSRM even improves the sampling efficiency over the fully random matrix and produce comparable performance with the state-of-the-art deep learning-based method without sacrificing the democracy. Additionally, RSRM also has the potential of reducing the bitrate overhead for CS-based IoT security systems.

## Acknowledgement

The authors would like to thank Dr. Trinh Van Chien for his thoughtful comments and Dr. Wuzhen Shi for making his source codes available for comparison. This work is supported in part by the National Research Foundation of Korea (NRF) grant (2017R1A2B2006518) funded by the Ministry of Science and ICT.